\documentstyle[twoside,fleqn,espcrc2,epsf]{article}

\newcommand{\AmS}{{\protect\the\textfont2
  A\kern-.1667em\lower.5ex\hbox{M}\kern-.125emS}}

\title{Monopoles and instantons in SU(2) lattice gauge theory}

\author{Tam\'as G.\ Kov\'acs\address{Department of Physics,
                University of Colorado,\\
                Boulder CO 80309-390, USA}
        and 
        Zsolt Schram\address{Department of Theoretical Physics, 
                Kossuth Lajos University,\\
                Debrecen H-4010, Hungary}}

\begin{document}
\begin{abstract}
We investigate the monopole-instanton
correlation in SU(2) lattice gauge theory
using a renormalisation group inspired smoothing
technique. 
We look at the properties of monopole 
clusters and their correlation with instantons. 
Since the action of the smoothed configurations is dominated
by instantons we compare the smoothed Monte Carlo lattices
to artificially reconstructed configurations 
with the same instanton content but no other fluctuations.
Both parallel and randomly rotated (in group space) instanton 
ensembles are considered.
\end{abstract}

\maketitle

\section{INTRODUCTION}

In SU(2) and SU(3) lattice gauge theories a strong correlation
has been found between Abelian monopoles and instantons \cite{all}.
In the earlier works either prepared instanton configurations
or cooled Monte Carlo configurations have been used to demonstrate
the connection between monopoles and instantons.
Lately renormalization group inspired methods were also
applied \cite{feu}.

As monopoles are generally believed to be
responsible for the non-Abelian confinement mechanism,
their correlation with instantons led to the assumption
that instantons might also play a role in confinement.
In the present work we investigate the monopole-instanton
correlation in SU(2) lattice gauge theory 
using a renormalisation group inspired smoothing
technique that was originally introduced to study instantons \cite{col1}.
The smoothing (which consists of an inverse blocking and a
subsequent blocking)
considerably reduces the noise without affecting
the long distance physics. We identify monopoles and instantons
on the lattice and
look at the properties of monopole 
clusters and their correlation with instantons. Different
configurations with different confining properties are used for
this purpose. In a further work we also consider
the monopole contribution to the string tension.

Simulations have been performed on an $8^3$x$16$ lattice using 
the FP  action of Ref.\ \cite{col1}. 30 configurations have been constructed 
via Monte Carlo method and
smoothed up to nine times. It has been shown in Ref.\ \cite{col1} that the SU(2)
string tension does not change during smoothing. Therefore, the
physical objects which are responsible for confinement (if there
are such objects) should be present on all of these configurations.

The topological charge density has been determined on all configurations.
On the 9 times smoothed configurations the size and position of instantons
and/or antiinstantons can be determined precisely. As the action on these
9 times smoothed  configurations is dominated by instantons, we compare
the smoothed Monte Carlo lattices to artificially reconstructed configurations
with the same instanton content but no other fluctuations. Both parallel and
randomly oriented (in group space) instanton ensembles are considered. 
The long distance 
properties of these artificial configurations are known as they are
identical with the configurations used in Ref.\ \cite{col2}. In particular, neither 
the parallel nor the randomly oriented instantons confine. 
Abelian monopoles have been identified 
both in the 
Maximally Abelian Gauge (MAG) and in the Polyakov Gauge (PG). However,
in the PG case not all the results are displayed here,
as
we believe that the effective Abelian model extracted in this gauge
has no consistent physical meaning.

\section{TOPOLOGICAL CHARGE DENSITY AND MONOPOLES}

In order to see 
the monopole-instanton correlation we calculate the average topological charge 
density on the monopole current links ($\langle q^2 \rangle_{mon}$) and 
the average charge on the whole lattice ($\langle q^2 \rangle_{tot}$).
If there is a correlation then one should find $\sqrt{\langle q^2 
\rangle_{mon} / \langle q^2 \rangle_{tot}} > 1$. Our results are shown
in the following table.

\vspace{1.5mm}
\begin{tabular}{|c||c|c|} \hline 
                   &{\small Maximal}&{\small Polyakov}\\ 
{\small configuration}&{\small Abelian Gauge}&{\small Gauge}\\ 
\hline \hline
 {\small original}   & 1.0969(5) & 1.002(2) \\ 
\hline
{\small 1 smoothing} & 1.91(4) & 1.13(2)  \\
\hline
{\small 2 smoothing} & 2.24(6) & 1.31(3)  \\
\hline
{\small 9 smoothing} & 3.28(15) & 1.91(8) \\
\hline
{\small random} & 4.25(20) & 2.04(7)  \\
\hline
{\small parallel} & 4.33(20) & 1.89(10) \\
\hline

\end{tabular}

\vspace{1.5mm}
The correlation is much stronger in MAG than in PG
at all stages of smoothing. Not unexpectedly, the correlation increases
with smoothing in both cases. We obtain, however, maximal correlation
for the artificial instanton configurations. The parallel and randomly
oriented instanton ensembles do not differ as far as the correlation is
concerned. Note that in PG the correlation is essentially the same for the
9 times smoothed configurations and for the artificial ones. In MAG,
in contrast, this is much higher for the prepared configurations than for any
Monte Carlo generated ones (up to 9 smoothing, of course). It is interesting
that the instanton configurations do not confine, and still, the monopole-instanton
correlation is the strongest for these, non-confining systems.

\section{MONOPOLE LOOPS AND INSTANTONS}

Quantities characterizing the monopole configurations are shown in an 
additional table. 
We observe that the first smoothing step removes roughly $90\%$ of the
monopoles. Still, the SU(2) string tension does not change. We have also
shown that the Abelian dominance found for the original, untouched
 configurations also weakens after one smoothing: the Abelian string tension
drops to $70\%$ \cite{abel}. Thus, if monopoles alone would provide the full string
tension of the effective Abelian model, then in MAG, $10\%$ of monopoles
should be responsible for most of the string tension. This is
in a qualitative agreement with the results of \cite{hart}.
\vspace{1mm}

\noindent
\begin{tabular}{|c||c|c|c|} \hline 
{\small  MAG}&{\small Number of}&{\small Number}&{\small Largest}\\ 
{\small config.}&{\small monopoles}&{\small of loops}&{\small loop length} \\ 
\hline \hline
{\small original}    &$1972(119)$&$77(10)$  & $1447(228)$ \\ 
\hline
{\small 1 smooth} & $219(36)$ & $5.5(2.0)$ & $122(40)$  \\
\hline
{\small 2 smooth} & $165(35)$ & $4.9(2.0)$ & $95(46)$  \\
\hline
{\small 9 smooth} & $83(27)$ & $4.3(1.5)$ & $41(21)$  \\
\hline
{\small random} & $43(18)$ & $5.6(2.0)$ & $15(9)$  \\
\hline
{\small parallel} & $41(16)$ & $6.1(2.1)$ & $11(7)$ \\
\hline
\end{tabular}  

\vspace{1mm}
The average instanton number for the configurations is $6.7\pm2.1$. The
average number of monopole loops for the smoothed configurations as well
as for the instanton ensembles, is approximately the same. Is there some 
explanation behind or this is simply an accident? 
We studied the number of monopole loops versus the instanton number
for all 
configurations. Correlation analysis of mathematical statistics
clearly showed a strong linear correlation between these quantities in case
of the
artificial 
instanton configurations. However, 
no correlation exists between the same quantities for the smoothed Monte
Carlo configurations.

This indicates the following. On the artificial instanton configurations
(both with parallel and random instanton orientation), in average, each
instanton is accompanied by a monopole loop close to its centre. The 
average instanton radius is $1.8\pm0.6$ and the instanton distribution
is dilute (see the average instanton number). The average length
of monopole loops is $6.8\pm 3.7$ for the parallel and $7.7\pm 5.3$ 
for the randomly oriented instanton ensembles, i.e.\ the loops are
small, local loops (see also the length of the largest loops in the table).
In contrast, for the smoothed configurations the average length  
of the monopole loops is $39.7\pm 44.8$ and $19.7 \pm 17.5$ for the once
smoothed and for the 9 times smoothed configurations, respectively.
On these configurations, large loops are formed which can cross several
instantons and the number of loops and number of instantons is uncorrelated.
We illustrate the situation for a configuration sample in Fig.\ 1 (which is 
available in colour from the hep/lat archives).
On the real configurations large loops are present which cannot be paired
with individual instantons. 
 
\section{CONCLUSIONS}

Using renormalization group inspired smoothing in SU(2) lattice
gauge theory we showed, that the first smoothing step drastically
reduces the number of Abelian 
      mono\-poles. Additional smoothing steps result only in moderate
      effect.

The correlation between monopoles and instantons increases with
      smoothing. We find a much stronger correlation for MAG than for PG.

In case of the artificially prepared instanton configurations 
      the size of the monopole loops is determined mainly by the size of 
      the instantons. The monopole-instanton correlation is the strongest 
      for these ensembles. In average, each instanton is accompanied by a 
      small monopole loop.
The Monte Carlo generated smoothed configurations differ considerably
      from the artificial ones. Large loops of monopoles are present. Work
      is in progress to determine the monopole contribution to the string
      tension at different stages of the smoothing sequence.

\section*{ACKNOWLEDGEMENTS}

We thank the Colorado Experimental High Energy 
Group and the UCLA Elementary Particle Theory group for 
granting us computer time. 
This work was partially supported by
U.S.\ Department of Energy grant
DE-FG02-92ER-40672, OTKA Hungarian Science Foundation T 23844 and
the Physics Research Group of the Hungarian Academy of Sciences,
Debrecen.

\vspace{-7mm}

\begin{figure}[htb]
\vspace{-0.72cm}
\begin{center}
\begin{minipage}{7.5cm}
\epsfxsize=4cm 
\epsfbox{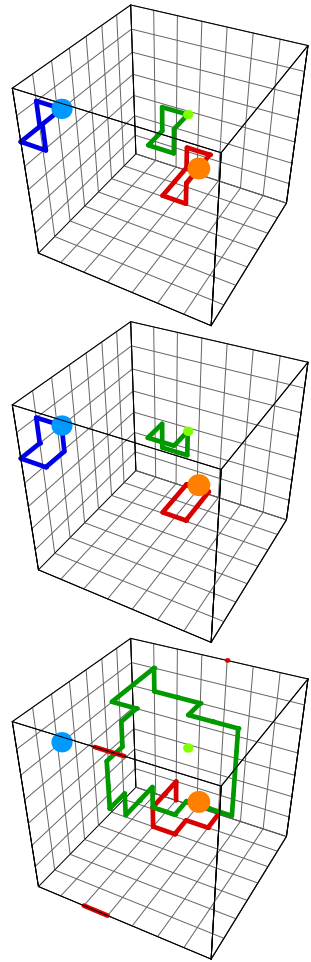}
\end{minipage}
\end{center}
\vspace{-10mm}

\caption{
Time projection of a configuration with 
three instantons for
artificial parallel (top) and random (middle) 
orientation
compared to the smoothed real 
one (bottom).
The instanton radii are 1.5$a$, 3.25$a$ and 3.5$a$
}
\label{fig:konf3}
\vspace{-0.5cm}
\end{figure}


\begin{thebibliography}{9}

\bibitem{all} 
      S.\ Thurner, H.\ Markum and W.\ Sakuler, hep-th/9506123;
      V.\ Bornyakov and G.\ Schierholz, Phys.\ Lett.\ B384 (1996) 190;
      A.\ Hart and M.\ Teper, Phys.\ Lett.\ B371 (1996) 261;
      H.\ Suganuma, A.\ Tanaka, S.\ Sasaki and O.\ Miyamura, Nucl.\
      Phys.\ B (Proc.\ Suppl) 47 (1996) 302
\bibitem{feu} 
      M.\ Feurstein, E.M.\ Ilgenfritz, H.\ Markum, M.\ M\"uller-Preussker
      and S.\ Thurner, hep-lat/9709140
\bibitem{col1} 
      T.\ DeGrand, A.\ Hasenfratz and T.G.\ Kov\'acs, Nucl.\ Phys.\ B505
      (1997) 417
\bibitem{col2} 
      T.\ DeGrand, A.\ Hasenfratz and T.G.\ Kov\'acs, hep-lat/9710078
\bibitem{abel} 
      T.G.\ Kov\'acs and Z.\ Schram, Phys.\ Rev.\ D56 (1997) 6824
\bibitem{hart} 
      A.\ Hart and M.\ Teper, Nucl.\ Phys.\ B (Proc.\ Suppl) 63 (1998) 522



\end{thebibliography}
\end{document}